\documentstyle[preprint,aps,epsfig,amssymb,float]{revtex}

\begin{document}

\title{Imprints of accretion on gravitational waves from black holes}

\author{Philippos Papadopoulos${}^{(1)}$ and Jos\'e A. Font${}^{(2)}$}

\address{(1) School of Computer Science and Mathematics,
University of Portsmouth,
PO1 2EG, Portsmouth, United Kingdom \\
(2) Max-Planck-Institut f\"ur Astrophysik,
Karl-Schwarzschild-Str. 1, D-85740 Garching, Germany}

\date{\today}

\maketitle

\begin{abstract}

Black holes are superb sources of gravitational wave signals, for
example when they are born in stellar collapse.  We explore the
subtleties that may emerge if mass accretion events increase
significantly the mass of the black hole during its gravitational wave
emission. We find the familiar damped-oscillatory radiative decay but
now both decay rate and frequencies are {\em modulated} by the mass
accretion rate. Any appreciable increase in the horizon mass during
emission reflects on the instantaneous signal frequency, which shows a
prominent negative branch in the $\dot{f}(f)$ evolution diagram. The
features of the frequency evolution pattern reveal key properties of
the accretion event, such as the total accreted mass and the accretion
rate. For slow accretion rates the frequency evolution follows
verbatim the accretion rate, as expected from dimensional
arguments. In view of the possibility of detection of black hole
``ringing'' by the upcoming gravitational wave experiments, the
deciphering of the late time frequency dynamics may provide direct
insight into otherwise obscured aspects of the black hole birth
process.

\vspace{0.5cm}

PACS numbers: 04.25.Dm, 04.40.-b, 04.70.Bw, 95.30.Lz

\end{abstract}

The classic black hole solutions of the Einstein equations are an
enduring achievement of general relativistic theory research as
evidenced by the increasing attention they receive in observational
programmes in astrophysics and gravitational wave (GRW) science.  {\em
Linearised perturbation} studies of such spacetimes are by now fairly
thoroughly understood (see~\cite{chandra,kokkotas}) and conclude that
the governing dynamical equations for perturbations of any spin
(scalar, electromagnetic and gravitational) are all mapped into wave
equations with a short range potential. Among the solutions of those
equations pre-eminent role is reserved for the damped oscillating
modes (out-going at infinity and in-going at the horizon), the
so-called quasi-normal modes (QNM). Physically, and in the language of
GRW signals, a generic excitation of an isolated black hole leads to a
prompt emission, followed by the exponentially decaying QNM
ringing. The latter is quickly dominated by the slowest damped mode.
A review of this phenomenology is given in~\cite{anderson}.  An
extension of the vacuum black hole paradigm is given
in~\cite{wai-mo}. This study concerned the computation of QNM
frequencies in the presence of matter shells, but is limited to
ad-hoc static spherical configurations.  Another class of simple
spherically symmetric solutions, those describing collapse of compact
lumps of matter, has also received wide
attention~\cite{oppenheimer,may-white}.  Extensively used in
conjunction with perturbation theory~\cite{cunningham,seidel}, those
solutions helped elucidate the GRW signals expected from stellar
collapse. These studies showed that the end-stages of the emission
during gravitational collapse may be dominated by the QNM ringing of
the emerging black hole, in particular the {\em single} QNM frequency
of the least damped mode. Early numerical relativity studies seem to
corroborate this picture~\cite{sp85}.

The precise dynamics of stellar collapse remains controversial,
especially for more massive stars. A range of so called ``collapsar''
models (see e.g.,~\cite{woosley} and references therein) speculate
that significant amounts of matter will accrete immediately {\em
after} the black hole has formed, possibly at very high accretion
rates. An expanding while radiating black hole is possible even in
complete absence of matter, e.g., in the coalescence of a binary black
hole. The paradigms of stationary vacuum black hole and simple
spherical collapse perturbations are perhaps inadequate models for GRW
emission in such cases, but the literature appears not to have
addressed the issue at all. In this work we initiate a study of this
regime using numerical, fully relativistic, solutions of accreting
black hole spacetimes and we model non-spherical spacetime
perturbations using the dynamics of a scalar field on this
time-dependent background. We find that accretion extends the
phenomenology of black hole ring-down in a well defined manner. The
resulting GRW signals are best described as amplitude and frequency
(AM/FM) modulated versions of the simple damped sinusoidal waveforms.

The first major component of the investigation is the self-consistent
relativistic description of a growing, spherically symmetric, black
hole spacetime. To our knowledge there exists no simple model for such
spacetimes, hence we resort to numerical simulation. As a sufficient,
not unduly specialised, model of an accreting black hole we solve the
coupled system of Einstein's field equations and perfect fluid
hydrodynamics in spherical symmetry as in~\cite{paper1}.  The
spacetime geometry is described in the Tamburino - Winicour
framework~\cite{winicour}, which provides a {\em characteristic
initial value formulation} for the system of Einstein's equations.
The computational domain is bounded by a worldtube $W$.  The metric
element is explicitly $ds^2 = - e^{2\beta}V/r d\tau^2 + 2 e^{2\beta}
d\tau dr + r^2 d\Omega^2 $, in geometrised units ($G=c=1$).  The
geometry is described by the two metric functions $\beta(\tau,r)$ and
$V(\tau,r)$. They obey radial hypersurface equations, with the
stress-energy tensor of the fluid acting as a source term. Our setup
uses lightcones converging to the interior of $W$ and intersecting the
black hole horizon. In the case of a static {\em vacuum} black hole,
this would correspond to advanced time (ingoing) Eddington-Finkelstein
coordinates. The metric data on $W$ are chosen to coincide with those
of the Schwarzschild metric at that radius ($\beta=0, V=1-2M/r$).  The
coupled evolution of the Einstein-matter system produces dynamical
spacetimes in the interior of $W$. The exterior region, initially
described by the static black hole solution, will remain described
consistently as such, provided that matter never reaches $W$. The
relativistic hydrodynamic equations for a perfect fluid are formulated
as a first-order flux-conservative hyperbolic system.  The fluid is
described completely by the rest-frame density $\rho$, the radial
velocity $u^{r}$ and the specific internal energy $\varepsilon$.  The
equation of state is $p=(\gamma-1)\rho \varepsilon$, with
$\gamma=5/3$.  The choice of fluid data on $\tau_0$ fixes the metric
functions $\beta$ and $V$ on that hypersurface. In turn those
characterise the nature of the spacetime (e.g., the presence and size
of a black hole).  For vacuum data (no fluid), an apparent horizon,
defined here as the spacelike two-sphere at which $V(\tau,r)$=0, will
be present on the initial slice at $r=2M$.  With non-zero fluid mass,
the initial location of the apparent horizon will shift towards
smaller radii. In our framework, the presence of an apparent horizon
is essential, as we rely on the concept of {\em excision} (See
discussion in~\cite{excision1}), as applied to the characteristic
initial value problem~\cite{excision2,excision3} for prescribing
boundary conditions at the inner edge of the domain.  Since the
apparent horizon must be at least at $r=0$, there is an upper bound to
the amount of matter that can be prescribed to the interior of $W$.
Further details of the implementation are given
in~\cite{paper1}. See~\cite{font} for recent reviews on techniques.

Whereas the previous considerations provide us with a fairly faithful
model of an accreting black hole spacetime, in order to identify the
features of gravitational wave emission we resort to more idealised
modelling. We capture the essence of non-spherical gravitational
perturbations by solving the evolution equation for a minimally
coupled mass-less scalar field $\Box\phi = (-g)^{-1/2} ((-g)^{1/2}
g^{\mu\nu} \phi_{,\mu})_{,\nu} = 0$.  The spacetime background on
which this equation is solved is the accreting black hole spacetime
provided by the Einstein-matter evolution system described above.  We
assume vanishing stress-energy density for the scalar field.
Physically, this setup closely models the dynamics of genuine
spacetime perturbations, and any differences are primarily numerical in
nature.  Characteristic initial data for the scalar field equation
consist of a single function $\phi_0(r)$ prescribed on the initial
ingoing lightcone $\tau_0$.  Such data simulate the initial {\em
shearing} of the in-going light-cone away from a pure spherically
symmetric convergence. Non-trivial deviations from sphericity
would be induced e.g., in the early phases of black hole formation due
to asymmetries in the collapse process. We make the simplifying
assumption that perturbations are not reinforced by any further
inhomogeneities during the accretion process, i.e., we treat a
decoupled and homogeneous problem for the wave equation. The linearity
of the scalar field and the spherically symmetric background allow the
expansion of $\phi$ into regular spherical harmonics
$Y_{lm}(\theta,\phi)$ and lead to a decoupled set of evolution
equations for each angular mode $g=r\phi_{l}(\tau,r)$: $ 2g_{,\tau r}
+ (V g_{,r}/r)_{,r} = (V/r)_{,r} g/r + l(l+1) e^{2\beta} g/r^2$.  The
integration procedure for this equation follows closely the techniques
used in~\cite{gomez1,gomez2,gomez3}. The algorithms are only slightly
modified by the presence of a horizon, inside which outgoing
characteristics converge.  The radial grid used in our simulations
extends from $r_{\mbox{min}}=0.6$ to $r_{\mbox{max}}=50$, with a
resolution $\Delta r= (r_{\mbox{max}}-r_{\mbox{min}})/1800$.  The
radial grid for the scalar field evolution coincides with the
spacetime grid for $r<r_{\mbox{max}}$ and extends three times as much
in the exterior, static region.  This latter step allows us to
effectively ignore the tricky question of what scalar field data to
prescribe on the worldtube so as to model the unimpeded transmission
of the wave into the static region.

We turn now our attention to the physical description of the accreting
black hole spacetimes. The ratio $\lambda=M_{\mbox{i}}/M_{\mbox{f}}$
of the initial to final black hole mass is a first key parameter.  For
a minimally complete parameter space, this number must be accompanied
by the average value of the accretion rate $\mu=<\!\!\dot{M}\!\!>$,
which for fixed $\lambda$ reflects the duration of the accretion
event.  We henceforth scale all distances and times (and derived
quantities) by the mass of the final black hole which is always
assumed to be unity $(M_{\mbox{f}}=1)$.  The collapse of a star to
form a black hole can be considered as the limiting case in which
$\lambda\rightarrow 0$.  In the opposite limit ($\lambda\rightarrow
1$) we deal with infinitesimal changes to the black hole mass.
Given the conventions above, the control of type of accretion event is
entirely through the choice of the fluid data on the initial time
surface.  It is intuitively obvious that only accretion events on mass
and time scales comparable to the size of the black hole horizon will
have an appreciable effect on the QNM emission. We select here a
family of fluid data that appears to be a fairly generic
representative of fast and significant accretion of matter. 

We choose an initial density profile that is radially constant up to a
radius $r_c$ and then decaying exponentially ($\rho_0(r) =
\rho_c$ for $r < r_c$ and $\rho_0(r) = \rho_c e^{- \kappa (r-r_c)^2}$
for $r > r_c$, with $\kappa=0.05$). The velocity and internal energy
profiles are simple monotonically decreasing power laws ($u^r\sim
r^{-0.5},\varepsilon \sim r^{-1}$). The initial density distribution
exerts the strongest influence on the evolution of the horizon, we
hence keep the other fluid variables fixed to the above choices and
vary the density profile via the parameters $\rho_c$ and $r_c$. We
turn next to the initial data for the scalar field. Physically, the
form (amplitude etc.) of the perturbation field at the initial time
depends heavily on the spacetime history {\em prior} to the event
simulated here.  We make the assumption that such perturbations have a
characteristic wavelength comparable to the size of the horizon at the
initial time $\tau_0$. We hence use $\phi_0(r) =
\phi_p$ for $r < r_p$ (peak) and $\phi_0(r) = \phi_p e^{-
\kappa (r-r_p)^2}$ for $r > r_p$. The amplitude $\phi_p$ is arbitrary.
Exploration of the parameter space $(r_p,\kappa)$ for the initial data
suggests that differences in the profiles affect mostly the early
phase of the signal (i.e., the prompt emission).  In the results
presented below we use $r_p = 1.06 M_{\mbox{i}}$ and $\kappa = 2$.

Figure 1 displays a spacetime diagram of a representative accretion
event.  The diagram focuses on the innermost region of the
computational domain, to help the identification of many of the key
elements of the simulation. We see the evolution of the apparent
horizon as a thick solid line. Representative streamlines (integrals
of the velocity vector field) of accreting matter are shown as thin
solid lines. The dotted lines foliating the spacetime diagram in Fig.1
form a central concept of the further analysis. They are the {\em zero
phase surfaces} of the perturbation field $\phi_l$.  At a remote
observer site those surfaces determine the zero crossings of the
signal.  The modulation of the period of the signal during the
depicted accretion process is evident already from the figure, as can
be seen from the variable intercept intervals, which grow longer at
later times.  In fact, the accretion is modulating both the decay rate
and the oscillation frequency, as shown in the typical signal profile
shown in Fig.3 Nevertheless, in the sequel we focus primarily on the
QNM instantaneous {\em frequency}, which clearly is the most relevant
aspect of the signal (a recent discussion of its relevance for GRW
detection was given recently in~\cite{putten}).  Instantaneous values
for the signal frequency (ISF) and its time derivative are obtained
from the zero crossings of the field measured by an observer located
at a fixed radius (here $r=30$).

In Figure 2 we show the main features of the ISF evolution in the form
of $\dot{f}(f)$ trajectories for different angular harmonic modes
($l=2,3,4$) and different accretion rates.  We see that all curves
share the same broad characteristics.  Initially, there is a rapid
sweep into negative values of $\dot{f}$. This reflects the fast
shedding of higher frequency (and higher damping rate) overtones from
the early emission, which results in a lower frequency signal.  The
evolution reaches a peak negative rate and then turns around and
asymptotes to the vacuum oscillation frequency ($\dot{f}$=0).
Different $l$ modes are seen to reach different negative rates and
seem to scale with the $l$ value.  This is a simple effect, due to the
change of scale of the frequency of the underlying signal.  We hence
explore the effects of varying accretion rate by fixing $l=3$.  We
present results for $r_c=(10,15,20)$ and peak densities $\rho_c\times
10^{5}=(0.69, 0.2752,0.1351)$, represented by empty diamonds, circles
and crosses, respectively.  All three choices correspond to
$\lambda=0.375$, i.e., a fixed ratio of initial to final black hole
mass.  The choice of $r_c$ effectively controls the duration of the
accretion event, albeit in a manner determined only {\it a
posteriori}. It is seen that larger $r_c$ values, that is, an initial
distribution of the accreting mass within a larger volume, lead to
larger variations of the frequency. The intuitive explanation of this
effect uses loosely notions from vacuum black hole perturbation
theory. A tenuous matter configuration implies a delayed growth for
the black hole horizon. In turn this means the QNM frequency samples
various ``horizon sizes''. A fast accretion, in turn, implies that
most of the ringing is produced around a black hole that has already
almost reached its final mass, hence the QNM frequency is mostly
around that final value.  The horizontal evolution of the QNM
frequency during the emission is then seen to reflect on the amount of
horizon growth during this era, whereas the elapsed time establishes
the accretion rate.  We'll see that for sufficiently slow accretion
rates this statement can be made very precise by using a simple
dimensional argument, but in the general case more detailed analysis
would be required to extract the precise correlation.  We note here
that Figure 2 illustrates in passing two tests of the numerical
procedure: i) A consistency test, represented by the vertical line at
$f=0.0769$.  This is the value for the QNM frequency of a vacuum black
hole (scalar $l=2$ mode), as estimated by perturbation theory (WKB
approaches)~\cite{iyer,kokkotas}.  ii) An accuracy test, represented
by the solid line overlaying the $l=4$ sequence, obtained with twice
the resolution in all parts of the algorithm.

In Figure 3 we show the late time behaviour of the negative branch in
the $\dot{f}(f)$ diagram, for a simulation with $l=3$ and $r_c=15$
(this behaviour is also generic to the family explored here).  The
solid line depicts the evolution of the mass accretion rate $\partial
log(M)/\partial\tau$ versus observer time $\tau$.  We identify as mass
$M$, the mass obtained from the radius of the apparent horizon
$M=r_{H}/2$, whose location can be read off by identifying the
location of the zero-crossing of the metric variable $g_{00}$.  At
late times the accretion flow slows to a trickle, and obeys a power
law slope (in this case the slope equals -4.5).  We find that the ISF
curve itself follows a power law of the same slope at late times.  For
a more thorough comparison, we may overlay the two data sets after
adopting a horizontal (time) offset for $\dot{M}(\tau)$ with respect
to $\dot{f}(\tau)$.  In the case shown this offset is $\tau_{O}=71$.
Perhaps not surprisingly, the best fit is produced by a value roughly
corresponding to the travel time of a null geodesic from the strong
field region around $r=3$ to the observer location at $r=30$. The good
agreement of the slopes shows that the specific dynamics of the black
hole growth in late times imprints itself rather directly on the ISF
and is hence communicated via gravitational waves to the exterior
domain. The QNM frequencies for a vacuum black hole of mass $M$ obeys
a scaling relation $\omega=\sigma/M$, where $\sigma$ depends only on
the overall shape of the potential black hole potential.  For
sufficiently small $\dot{\sigma}$ we obtain that the logarithmic
derivative of the frequency equals the negative logarithmic accretion
rate ($\partial log(f)/\partial \tau=-\partial log(M)/\partial \tau$).
We have seen in Figure 3 that indeed this behaviour is dominant in
late times, suggesting indeed that $\dot{\sigma}$ is small in this
scenario.

It may be useful to try to establish closer connections with vacuum
black hole perturbation theory.  By introducing a coordinate $x(r)$
(generalising the {\em tortoise} coordinate~\cite{chandra}) governed
by $dx/dr=r/V$, the wave equation for a scalar field on a growing
black hole background reads $ 2g_{,x\tau}+ C g_{,xx} + D g_{,r} + P g
=0 $ where $C=1 + 2dx/d\tau$, $D=-V_{,\tau}/r$ and $P=-V/r
[(V/r)_{,r}/r + l(l+1)e^{2\beta}/r^2] $.  This equation generalises
analogous expressions for static black holes.  The function
$P(\tau,r)$ may be thought of as an ``instantaneous'' effective
potential.  The dashed line in Figure 1 illustrates the outward motion
of the maximum of this function.  The trajectory $r_P(\tau)$ follows
quite closely the horizon motion, and interestingly, the ratio
$r_P(\tau)/r_H(\tau)$ is found to be almost constant as a function of
time, irrespective of the accretion rate (hence equal to the value of
that ratio for a {\it static} black hole).  This is somewhat
surprising, as it implies that $\dot{\sigma}$ is almost zero.  While
this is consistent with the late time agreement of Fig.3, we see that
at early times the frequency evolution is visibly different from the
mass evolution.  We conclude then that at early times (and rapid
accretion), the dynamics is more subtle than the one deduced from
dimensional arguments, but potentially explainable through the
additional terms in the generalised wave equation introduced above.

Part of this work was completed during a visit of J.A.F to Portsmouth,
supported through PPARC and MPA. Computations were performed on a
Compaq XP1000 workstation at the University of Portsmouth. We warmly
thank N.~Anderson and K.~Kokkotas for comments on the
manuscript. J.A.F is grateful to the Relativity and Cosmology Group at
the University of Portsmouth for generous hospitality.

\begin{figure}[t]
\vspace{-0.5cm}
\centerline{\epsfig{figure=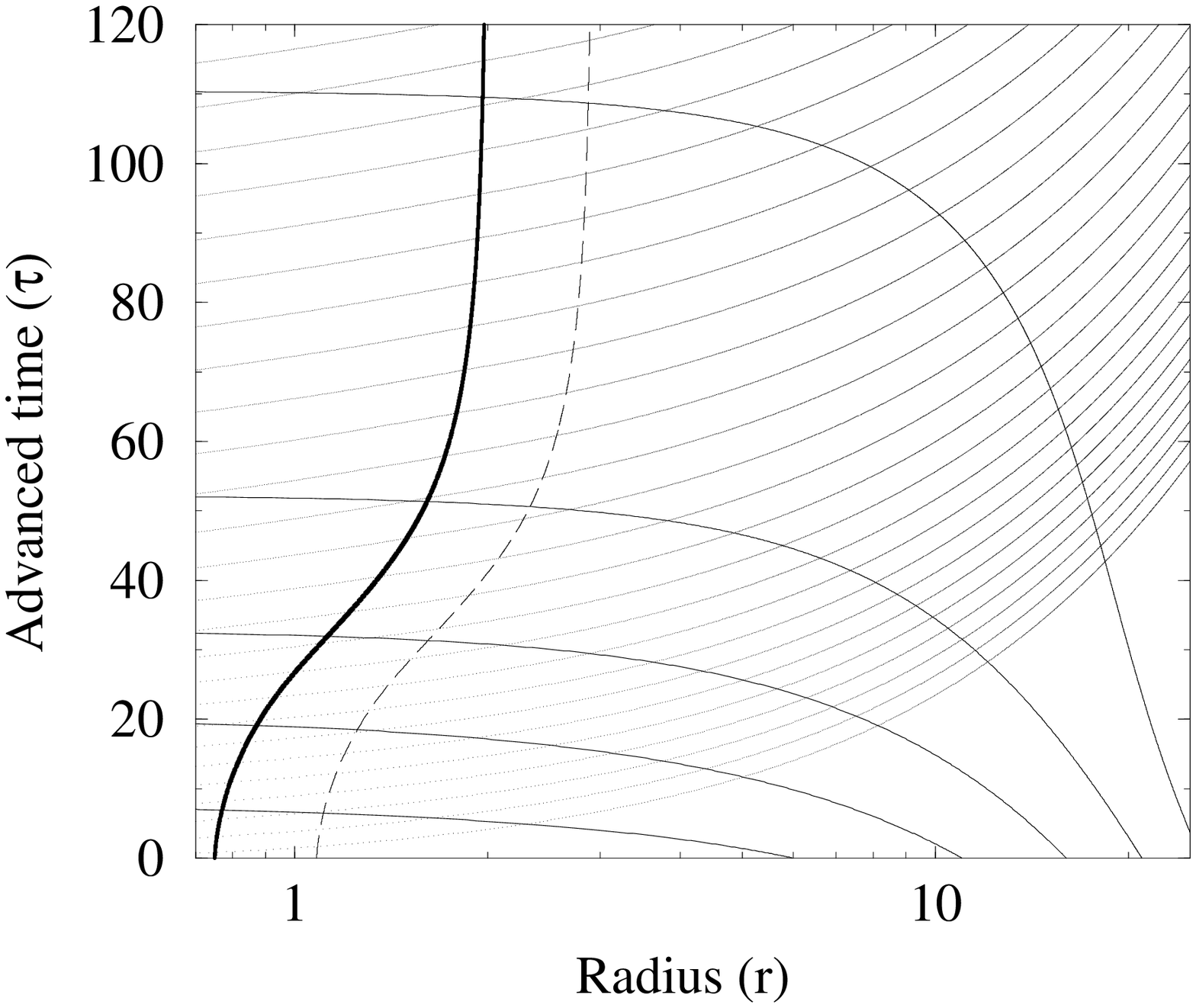,height=12cm,width=14cm}}
\caption{ Scalar field dynamics on an 
accreting black hole spacetime.  The thick solid line is the
trajectory $r_H(\tau)$ of the apparent horizon, starting from the
initial value $r=0.75$ and growing to the final value $r=2$.  The
infall of fluid matter is illustrated by representative streamlines,
shown as thin solid lines. The lines start at initial radii $r_{i} = 6
+ (i-1) 5$, with $i=1,\ldots,5$.  The dotted lines foliating the
diagram are the zero phase surfaces of the perturbation field
$\phi_l$.  For any given field multipole, the diagram represents the
appearance of the gravitational wave generation and propagation for a
fixed set of angular directions $(\theta,\phi)$.  The zero phase
surfaces tend asymptotically to outgoing light cones (the use of a
logarithmic plot obscures this somewhat in the diagram), but are
clearly spacelike near the horizon.  It is readily seen that an
observer located at a constant radius will intercept zero phase
signals of initially shorter periods which progressively increase and
asymptote to the QNM period of the final black hole.  Tracing back the
zero phase surface to the horizon region illustrates how the size of
the horizon is associated with the temporal separation of successive
levels.}
\label{spacetime}
\end{figure}

\begin{figure}[t]
\vspace{-0.5cm}
\centerline{\epsfig{figure=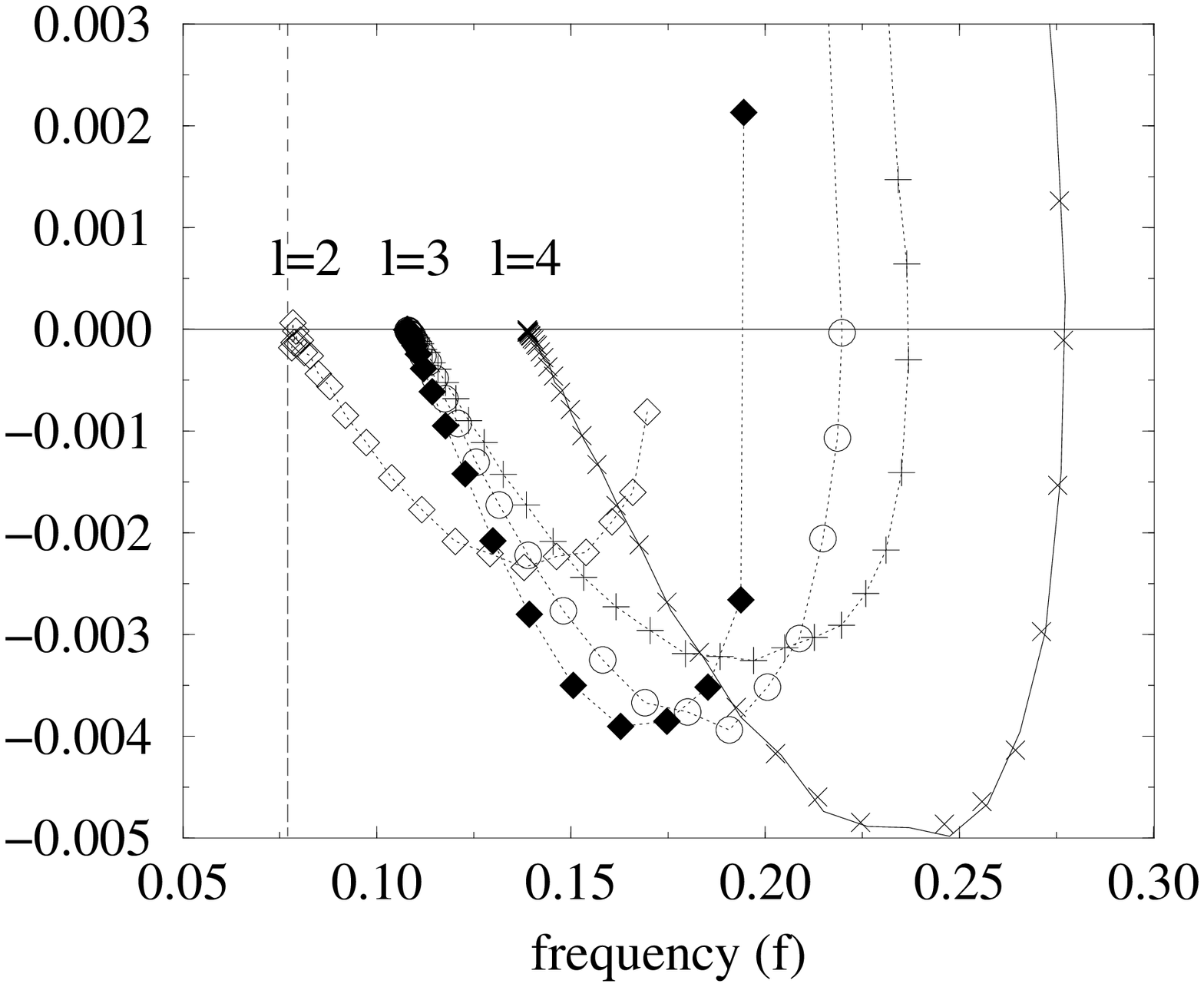,height=14cm,width=15cm}}
\caption{Frequency time derivative versus frequency.
The symbols (crosses, circles etc.) denote instances at which a zero
crossing event enables the computation of an instantaneous frequency
and its time derivative. The fitted lines aid the reading of the
trajectories.  Common to all evolutions is the rapid sweep into
negative $\dot{f}$ in the early phase of the signal. The evolution
reaches its peak (negative) $\dot{f}$ and as the accretion ceases
asymptotes to the vacuum oscillation frequency
($\dot{f}$=0). Different $l$ modes provide qualitatively similar
probes of the frequency dynamics, but with an overall magnitude
difference. The vertical dashed line at $f=0.0769$ denotes the
frequency of a scalar ($l=2$) {\em vacuum} black oscillation (of the
same final mass), as estimated by a higher order WKB
approximation. The solid line overlaying the $l=4$ sequence is
obtained with double resolution in all parts of the algorithm and
indicates that the resolution used is sufficient.}
\label{fdot}
\end{figure}

\vspace{-0.5cm}
\begin{figure}[t]
\centerline{\epsfig{figure=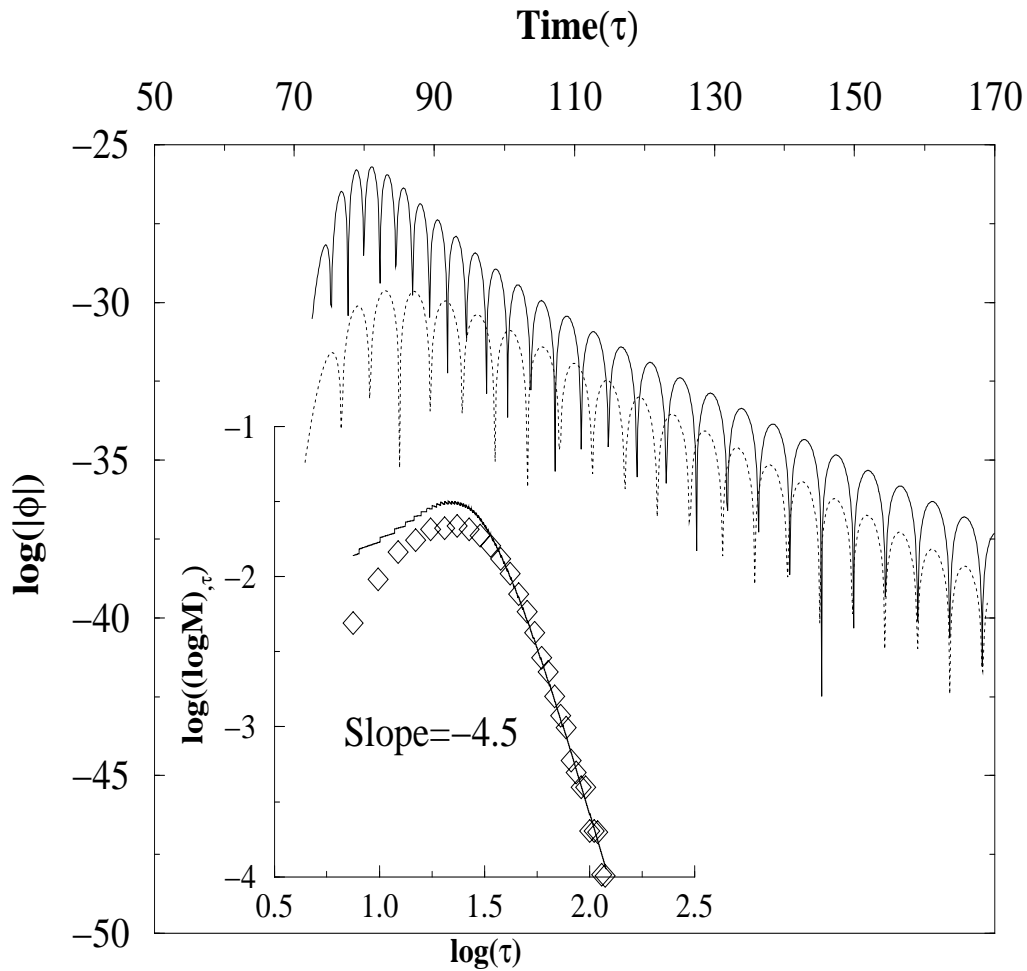,height=14cm,width=15cm}}
\caption{Typical modulated waveforms and correlations between 
frequency and accretion rates. The logarithm of the signal is shown as
a function of time for an $l=3$ and $r_c=15$ simulation (solid
line). The dotted line indicates the ringdown of a {\em vacuum} black
hole with $M=1$. The decay rate modulation is particularly evident
here. We have not analysed this effect quantitatively.  The insert
shows the late time behaviour of the signal frequency in correlation
with the accretion rate, as a function of time (same simulation) The
solid line depicts the evolution of the accretion rate $dlog(M)/d\tau$
versus observer time $\tau$. That quantity is derived from the
location of the horizon and is governed directly by the amount of
inflowing fluid.  Overlayed on the mass accretion rate is the
logarithmic time derivative of the signal frequency.  At late times
the accretion flow leads to a power law slope (in this case the slope
equals -4.5) and we find that the ISF curve itself follows a power law
of the same slope.}
\end{figure}

\end{document}